\documentclass[pra,aps,twocolumn,notitlepage,superscriptaddress,showpacs,nofootinbib]{revtex4-2}
\usepackage{enumerate,appendix}
\usepackage{amsmath,amssymb,amsthm,amsfonts,mathrsfs}
\usepackage{mathtools}
\usepackage{bbold} 
\usepackage{optidef}

\usepackage{hyperref}
\usepackage{stmaryrd}

\usepackage{bbm}

\usepackage{graphicx,color,cases} 
\usepackage{xfrac}  
\definecolor{DarkGreen}{rgb}{0.1,0.5,0.1}
\definecolor{DarkRed}{rgb}{0.5,0.1,0.1}
\definecolor{DarkBlue}{rgb}{0.1,0.1,0.5}

\allowdisplaybreaks

\theoremstyle{definition}

\newtheorem{remark}{Remark}
\numberwithin{equation}{section}
 
\def\>{\rangle} 
\def\<{\langle}

\newcommand{\ket}[1]{\left\lvert #1 \right\rangle}
\newcommand{\bra}[1]{\left\langle #1 \right\rvert}
\NewDocumentCommand\ketbra{+m+g}{%
  \IfNoValueTF{#2}
    {\left\lvert #1 \right\rangle \left\langle #1 \right\vert}
  {\left\lvert #1 \right\rangle \left\langle #2 \right\rvert}%
}
\NewDocumentCommand\braket{+m+g}{%
  \IfNoValueTF{#2}
    {\left\langle #1 \vert #1 \right\rangle}
  {\left\langle #1 \vert #2 \right\rangle}%
}

\newif\ifcomment
\commenttrue

\usepackage{quantikz}
\begin{document}

\title{GHZ is All You Need: Quantum Sensing with VISTA}

\author{Oskar Novak}
\email{onovak@arizona.edu}
\affiliation{Department of Electrical and Computer Engineering, The University of Arizona, Tucson, Arizona, 85721, United States of America}
\author{Christos N. Gagatsos}
\email{cgagatsos@arizona.edu}
\affiliation{Department of Electrical and Computer Engineering, The University of Arizona, Tucson, Arizona, 85721, United States of America}
\affiliation{Wyant College of Optical Sciences, The University of Arizona, Tucson, Arizona, 85721, United States of America}
\affiliation{Program in Applied Mathematics, The University of Arizona, Tucson, Arizona 85721, United States of America}
\author{Narayanan Rengaswamy}
\email{narayananr@arizona.edu}
\affiliation{Department of Electrical and Computer Engineering, The University of Arizona, Tucson, Arizona, 85721, United States of America}

\begin{abstract}
Quantum metrology holds the potential to enhance magnetic field sensing beyond current limits. However, in the presence of realistic noise, this advantage degrades to the Standard Quantum Limit. While recent algorithmic and variational techniques attempt to recover this scaling, they are hindered by stringent control requirements on the probe state that are infeasible in the near term, or by barren plateaus and interpretability issues inherent to black-box variational quantum circuits. Here, we introduce Variational Inference and Sensing with Twin Ans\"atze (VISTA), a closed-loop protocol that combines passive sensing, or where the probe state is left to evolve without any active control, with physics-informed variational optimization. In the VISTA framework, a probe state evolves under a Lindbladian master-equation, and is compared, via the Swap test, to a parameterized ``quantum twin", a shallow quantum circuit designed to mimic the underlying pure-state or Lindbladian master-equation dynamics. By restricting the optimization space to the physical parameters of interest, VISTA  circumvents barren plateaus. We demonstrate that by coupling the protocol with a classical optimizer and high shot counts, VISTA can temporarily achieve near-Heisenberg scaling for moderately noisy qubits over a finite range of system sizes. Furthermore, we introduce a Quasi-Normalization technique that sharpens the loss gradients, enabling simultaneous extraction of both the coherent signal $\theta$ and the environmental noise rate $\gamma$ with low absolute error. Finally, we extend VISTA to the multi-parameter vector metrology regime, enabling simultaneous parameter extraction from a transverse-magnetic-field Hamiltonian. By eliminating the need for complex, open-loop control and processing, VISTA offers a highly practical, resource-efficient framework for near- to intermediate-term quantum sensors.
\end{abstract}

\maketitle

\section{Introduction}
\label{sec:intro}

Quantum metrology is a key technology for the second quantum revolution, as it promises to estimate parameters that may be difficult to measure with high precision using current classical approaches. However, the traditional approach of using an entangled probe state to increase sensitivity and performing active error correction to fend off environmental noise cannot exceed the Standard Quantum Limit $\frac{1}{\sqrt{N}}$ in many realistic settings~\cite{Zhou-natcomm18}. This renders the quantum sensing advantage irrecoverable asymptotically for noise channels relevant to metrology, such as the dephasing channel. 

To circumvent this result, a recent surge of work has focused on using quantum and classical computation to try to regain a metrology advantage. Approaches range from black box variational quantum circuits, machine learning to find more sensitive probe states or measurement schemes \cite{le2023variational,meyer2021variational,koczor2020variational}, and quantum algorithms such as Grover's search or quantum signal processing as part of the sensing loop to either enhance AC (alternating current) sensing or reduce errors to achieve Heisenberg Scaling \cite{allen2025quantumcomputingenhancedsensing,marrero2026encodedquantumsignalprocessing}. Other approaches use black-box digital-twin reinforcement learning to find optimal control strategies for a probe state \cite{xu2025learningrestoreheisenberglimit}, and a similar approach that uses autonomous QEC \cite{kwon2026restoringheisenbergscalingtime}. Though many of these approaches do recover the Heisenberg Limit, they all require ultra-precise, real-time control over the probe state and measurement, which is difficult to achieve in practice, especially as we enter the early fault-tolerant era. Also, variational approaches typically employ black-box or uninterpretable quantum circuits and models, which not only preclude a useful interpretation of the learned state, but also suffer from vanishing gradients for variational ansätze that require too many parameters and entanglement \cite{Larocca_2025}. Finally, Grover's search and quantum signal processing are notoriously sensitive to noise, rendering these approaches effective only for fault-tolerant quantum computers \cite{Wang_2020,Escudero_2023}. 

In this work, we propose an alternative approach that circumvents all of these issues, which we call Variational Inference and Sensing with Twin Ansätze (VISTA)---a closed-loop protocol that merges the passive sensing paradigm of traditional metrology with variational optimization. In the VISTA framework, a simple $N$-qubit GHZ probe state passively evolves in a noisy magnetic field for a fixed duration. Then, we teleport the state to a register in a quantum computer, where a ``quantum twin" has been prepared while the probe state was evolving. This twin is a shallow, \emph{physically-motivated}, variational quantum circuit that explicitly mimics the expected pure or Lindbladian master-equation dynamics of the probe state evolution. The output state of this circuit is compared to the state of the probe via the Swap test to calculate the Hilbert-Schmidt inner product between them, which we use as part of a loss function to learn our signal parameter $\theta$. Because the circuit ansätz directly models the physical evolution, VISTA circumvents the barren plateau problem, reducing the optimization space to just the parameters of interest. With a sufficient number of swap test shots per epoch, we show that VISTA, when coupled with a momentum-based classical optimizer such as ADAM, transcends the limitations of traditional independent sampling. 
Iterative optimization accumulates information about our parameters over time, resolving the curvature of the loss landscape to achieve low absolute error in our estimation of $\theta$. By using a high shot count, we show that it is possible to achieve near-Heisenberg scaling for moderately noisy qubits over a finite range of $N$ values. Although this scaling may not last in the asymptotic $N$ limit, the absolute error of $\theta$ is low enough in this region to enable an accuracy of $\theta$ that would require significantly more quantum resources for other approaches. 

We can extend this protocol by using variational quantum ansätze that simulate the Lindbladian dynamics to learn both $\theta$ and the channel noise rate $\gamma$. To prevent the gradients from vanishing here, we introduce a new technique called \textit{Quasi-Normalization}, which allows for good convergence even when the ansätz is noisy. Finally, we briefly discuss generalizations of VISTA to the multi-parameter vector metrology regime, in which the Hamiltonian has a transverse component and an additional signal parameter $\theta_{2}$. The simplicity of our protocol's quantum circuits, and the robustness of our approach to noise due to the use of the Stochastic Parameter Shift Rule \cite{Banchi_2021}, and ADAM, make our approach a suitable application for near-term to intermediate-term quantum computers. We note a similar approach in \cite{wang2025noiseresilientquantummetrologyquantum}, which also relies on passive probe state evolution and state transfer to a quantum computer. However, VISTA offers several distinct advantages. Instead of relying on quantum principal component analysis (QPCA) via a variational density-operator exponentiation scheme to purify the probe state, VISTA uses a simpler, highly interpretable variational twin circuit. Because VISTA models the Lindbladian dynamics rather than merely filtering the noise, it allows us to learn the environmental noise rates alongside the signal. Furthermore, this simulation-based approach naturally accommodates non-commuting observables, providing a seamless extension to vector metrology that is fundamentally difficult to achieve with purely QPCA-based state purification. 

The paper is organized as follows: In Section~\ref{sec:backroundreview}, we give a brief review of quantum metrology. In Section~\ref{sec:protocoloverview}, we detail the VISTA protocol and its performance in estimating $\theta$, and introduce a cascading approach for scaling to large $N$. Section~\ref{sec:tradition} provides a direct numerical comparison between VISTA and a traditional quantum metrology protocol. In Section~\ref{sec:Noisycircansätz}, we construct explicit noisy circuit ansätze for dephasing and amplitude damping channels, and remark on how to generalize this approach to more exotic channels.  In Section~\ref{sec:performance noisy}, we discuss the analytical performance bounds for our noisy circuit ansätze for estimating $\theta$, compare them to the pure case, and introduce Quasi-Normalization to help converge while learning both $\theta$ and $\gamma$ without too much more error in $\theta$. In Section~\ref{sec:multiparam}, we briefly discuss the generalization of VISTA to the multi-parameter vector metrology regime. 
Finally, in Section~\ref{sec:discussion} we conclude by discussing the outlook given our results.

\section{Review of Quantum Metrology}
\label{sec:backroundreview}

Given a probe state $\rho$ evolving under a Hamiltonian of the following form, scaled by parameter $\theta$:
\begin{equation}
    \label{hamiltonian}
    \hat{H}=\sum_{j=1}^N \hat{Z}_{j},
\end{equation}
we wish to estimate $\theta$ with maximum precision. This is bounded by the Cram\'er-Rao bound \cite{Braunstein-prl94}:
\begin{equation}
    \label{CRB}
    \delta \theta \geq \frac{1}{\sqrt{\mathcal{Q}(\rho_{\theta})}}, 
\end{equation}
where $\mathcal{Q}(\rho_{\theta})$ is the Quantum Fisher Information (QFI) \cite{paris2009quantum}:
\begin{equation}
    \label{QFIintro}
    \mathcal{Q}(\rho_{\theta})=2\sum_{i\neq j}\frac{|\bra{i}\frac{\partial \rho_{\theta}}{\partial \theta}\ket{j}|^{2}}{p_{i}+p_{j}},
\end{equation}
given that the probe state $\rho$ is written in its spectral decomposition $\rho=\sum_{i}p_{i}\ket{i}\bra{i}$. Typically, the limits of interest for \eqref{CRB} are the Heisenberg Limit (HL), $\delta\theta \geq \frac{1}{N}$, and the Standard Quantum Limit (SQL),  $\delta\theta \geq \frac{1}{\sqrt{N}}$, where $N$ is the number of qubits. It is typically the goal of quantum metrology to reach the HL or at least exceed the SQL, but this becomes challenging in the presence of noise, even with active error-correction approaches \cite{Zhou-natcomm18}.

\section{Overview of the Metrology Protocol}
\label{sec:protocoloverview}

VISTA takes a unique approach to quantum metrology. Instead of relying on a sophisticated probe state or active control of one, we use an $N$-qubit GHZ state: 
\begin{equation}
    \label{ghz}
    \ket{\textrm{GHZ}} \coloneqq \frac{1}{\sqrt{2}}\ket{0}^{\otimes N}+\frac{1}{\sqrt{2}}\ket{1}^{\otimes N},
\end{equation}
which is allowed to evolve passively for a fixed duration in the presence of a magnetic field. This evolution is characterized by a Lindbladian master-equation \cite{manzano2020short}:
\begin{align}
    \label{Linbladian}
    \frac{\partial \rho}{\partial t}=-\frac{i}{\hbar}  [\hat{H},\rho]\theta+\sum_{k}\left(L_{k}\rho L_{k}^{ \dagger}-\frac{1}{2}\{L_{k}^{\dagger}L_{k},\rho\}\right),
\end{align}
where the Hamiltonian is given by \eqref{hamiltonian}, and $\rho=\ket{GHZ}\bra{GHZ}$. The quantum state of the probe is then transferred to the register of a quantum computer. As the probe was evolving, another $N$-qubit GHZ state was prepared and evolved by a quantum circuit on that same quantum computer, where two different ansätze may be used: a pure state variational ansätz where the GHZ state only undergoes Hamiltonian evolution, and a noisy ansätz, where a circuit that emulates the Lindbladian evolution is used instead to also learn the noise rate $\gamma$. Once both the probe state and the circuit ansätz have been loaded into the quantum computer's registers, we compute the Hilbert-Schmidt inner product via the swap test. This procedure is typically repeated for $ 10k$ to $40k$ shots, but more shots will yield better performance. We use $\mathcal{L}=1-\mathrm{Tr}(\rho_{probe}\rho_{circ})$ as our loss function.  

\begin{figure}[!t] 
    \centering
   \includegraphics[scale=0.6, keepaspectratio]{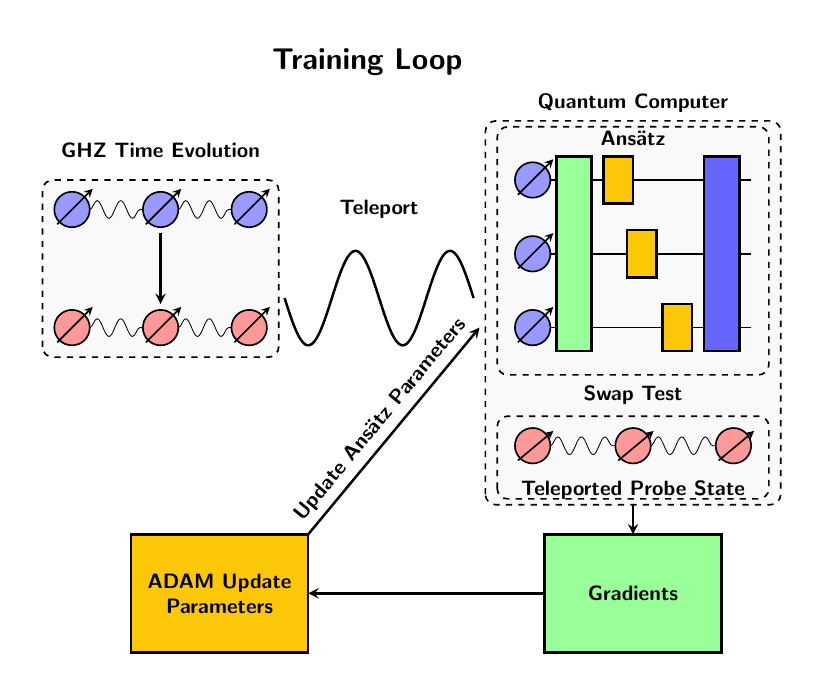}
    \caption{A diagram summarizing the VISTA metrology protocol. First, the probe state is prepared and allowed to passively evolve under the influence of the magnetic field and noise of its surroundings. The quantum state of the probe is then teleported to the registers of a quantum computer, while the ansätz state evolves, thereby mimicking the probe state's evolution using a variational quantum circuit that acts as a digital twin. The quantum computer then performs the swap test over multiple shots, each with a fresh probe state, to compute the Hilbert-Schmidt Inner Product between the circuit ansätz and the probe state. Gradients for this quantity are computed using the Stochastic Parameter Shift Rule and passed to the classical optimizer ADAM, which updates the circuit parameters for the next epoch. This repeats until convergence, i.e. when the parameters do not update significantly.}
    \label{maindiagram}
\end{figure}

In the noisy ansätz case, we use a technique called \textit{Quasi-Normalization} to sharpen the gradients when learning both the signal and the noise. We discuss this more in Section~\ref{sec:performance noisy}. Then, we use the Stochastic Parameter Shift Rule \cite{Banchi_2021} to calculate the gradients. This technique allows us to calculate gradients that are robust to the shot noise from the swap test, which becomes our main source of error in the protocol on fault-tolerant hardware. Next, we pass these gradients to ADAM \cite{kingma2017adammethodstochasticoptimization}, which then updates our guess for $\theta$ and $\gamma$ respectively. Since we are using circuit ansätze modeled on the system's physical evolution, \emph{we do not need to use a typical black-box variational ansätz to learn our parameters}. Instead, for the case of a single-parameter metrology problem where we also learn the noise rate of a dephasing channel, we only need to learn 2 parameters, allowing us to circumvent the typical pitfalls of a vanishing gradient for variational quantum circuits \cite{Larocca_2025}, and allowing us to have an interpretable model. This protocol simplifies the typical metrology process by making the primary metrological advantage come from the number of shots used during the swap test, and ADAM's ability to converge over noisy gradients. This is due to ADAM accounting for the shallowness of the loss landscape and its exponentially weighted moving average (momentum) term, which keeps the solver from oscillating near the minimum \cite{kingma2017adammethodstochasticoptimization}. We trade off the complexity of controlling and preparing a novel probe state, as in typical protocols, for computational complexity in terms of the number of epochs and shots per epoch. This allows us to use far simpler probe states. GHZ is selected because it is one of the simplest entangled probe states to prepare, and in the pure case, saturates the Cram\'er-Rao bound \cite{11396347}. We summarize our protocol in Fig~\ref{maindiagram}.
\label{sec:protocol}

\subsection{Discussion of Numerical Simulation Techniques}

We employ the ITensors.jl library \cite{fishman2022itensor} to simulate the quantum circuit using the Time Evolving Block Decimation (TEBD) algorithm \cite{PhysRevLett.91.147902} on an MPS tensor-network, and the Time-Dependent Variational Principle (TDVP) algorithm to simulate the probe state's vectorized Lindblad evolution \cite{PhysRevLett.107.070601}. Both networks are converted into doubled tensor networks in a Liouville space \cite{westhoff2025tensornetworkframeworklindbladian}, allowing the unnormalized inner product between them to be calculated. We feed this exact value to a binomial distribution sampler to emulate the shot noise from the swap test. From here, we use the sampler's output values to calculate the gradients. Since the GHZ state for any $N$ has a fixed bond dimension of $\chi=2$, and the Lindbladians considered for a noisy magnetic field have a non-entangling Hamiltonian, it is possible to simulate the dynamics of both the circuit and probe state in $O(N)$ time, respectively. Finally, to help control the convergence, we employ an exponential learning rate schedule to help with the optimizer reach the minimum without overshooting.

\subsection{Protocol Performance for Learning $\theta$}

\begin{figure}[!t] 
    \centering
    \includegraphics[width=0.5\textwidth, height=6cm, keepaspectratio]{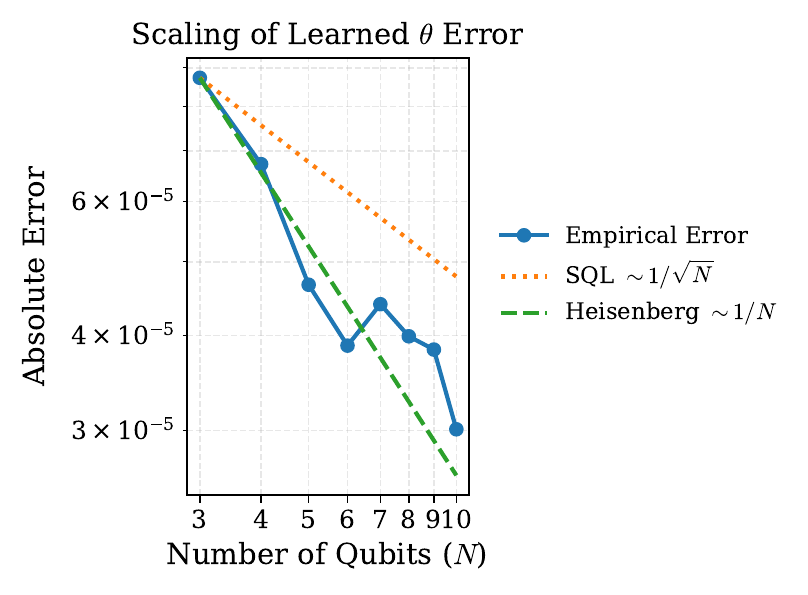} 
    \caption{For a moderate noise rate of $\gamma=0.005$, and $\theta=0.05$, we can achieve near-Heisenberg scaling by employing a shot budget of $10^{5}$ shots. This gives a scaling law of $\delta \theta \approx N^{-0.88}$, but this scaling can be improved to $N^{-1}$ with more shots. However, in the asymptotic limit as $N$ increases, the scaling is expected to converge towards the SQL due to noise.  Regardless of the large-$N$ behavior, this demonstrates that the VISTA protocol, by leveraging shots to stabilize the gradients of a variational quantum algorithm, can achieve a clear metrological advantage in certain regions. Note that the empirical error dips beneath Heisenberg scaling due to variance in the optimizer convergence.}
    \label{heisenberg}
\end{figure}

We see from \eqref{crunnorm} that for the pure ans{\"a}tz case, the Cram\'er-Rao bound from the loss function for dephasing is:
\begin{equation}
\label{crupure}
    \delta \theta \geq \frac{e^{N\gamma}}{2\sqrt{\nu}N},
\end{equation}
where $\nu$ is the number of shots taken during the swap test \cite{PhysRevLett.87.167902}. We see that for a reasonable $\gamma$, or noise rate, we can nearly recover HL up to a certain point, when $\nu$ is sufficiently high to stabilize the gradients. We see this in Fig~\ref{heisenberg}, where we also observe low absolute errors for each point, stemming from the number of shots that help resolve the curvature of the loss function near the minimum and from ADAM's ability to reach it. Although this scaling behavior arises from using a high number of shots, it demonstrates a distinct operational advantage of the VISTA protocol: it can efficiently leverage a larger classical sampling budget to increase the precision of quantum measurements. 

To understand this advantage, consider a typical protocol such as Ramsey fringe interferometry \cite{ramsey1950molecular}. In a standard Ramsey experiment, using a similar number of shots yields a lower-variance average estimate of a parity operator at a given time t, and these measurements at each time step are independent of one another. In contrast, the optimization process here is iterative: the parameter values from the previous epoch serve as the starting point for the next. This distinction means the shots compound information about $\theta$ over time by producing more stable gradient values. As the loss landscape steepens with increasing $N$, the stabilized gradients allow the optimizer to better navigate toward values significantly closer to the actual $\theta$. This advantage of using a gradient optimizer is discussed further in Section~\ref{sec:tradition}, where we compare the VISTA protocol to a traditional metrology protocol for a GHZ state.

\begin{figure}[!t] 
    \centering
    \includegraphics[width=0.5\textwidth, height=5cm, keepaspectratio]{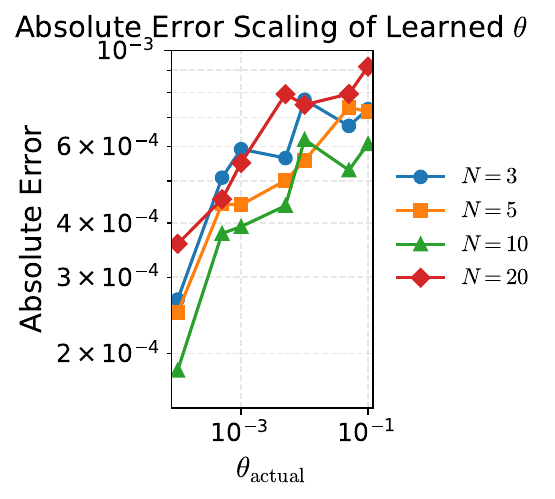} 
    \caption{Averaged Absolute error for a typical range of $\theta$ values for different-sized GHZ states with a more aggressive noise rate of $\gamma=0.05$. We use a shot scheduler ranging from $10$K to $40$K shots per epoch.  We notice that due to the higher noise rate, there are diminishing performance returns past $N=10$, with an $N=20$ GHZ state performing worse than an $N=3$ GHZ state.}
    \label{abserrortheta}
\end{figure}

However, from \eqref{crupure} we also notice that for sufficiently high $N$, there will be diminishing returns in performance due to the noise rate causing signal decay. This will lead to HL scaling beginning to vanish, even if the absolute error of $\theta$ is still low, as shown in Fig~\ref{abserrortheta}.
Here, we see that for this range of $\theta$ values, we have a diminishing metrological advantage after $N=20$ qubits for an aggressive noise rate of $\gamma=0.05$. But for more moderate noise rates, we can continue to scale up the number of qubits further until we start to see similar behavior.

\subsection{Scaling to Larger $N$}

\begin{figure}[!t] 
    \centering
   \includegraphics[scale=0.5,keepaspectratio]{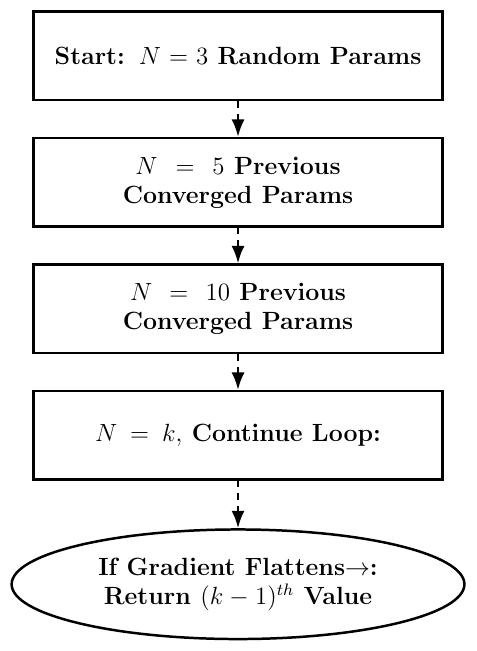}
    \caption{A diagram summarizing the cascade approach for scaling to large $N$.}
    \label{cascade}
\end{figure}

In cases where the noise rate $\gamma$ is sufficiently small that larger $N$ probe states yield better precision, one must be careful to ensure that the initial guess is within the correct convexity region of the loss function. We see this by looking at the case with two pure $N$-qubit GHZ states, $\ket{\theta_{1}}$, $\ket{\theta_{2}}$, where:
\begin{equation}
\begin{split}
    \label{pure evolution}
   \ket{\theta_{1}}=\frac{1}{\sqrt{2}}\left(e^{-\frac{i}{\hbar}N\theta_{1}}\ket{0}^{\otimes N}+e^{\frac{i}{\hbar}N\theta_{1}}\ket{1}^{\otimes N}\right),
\end{split}
\end{equation}
and the fidelity between the two states is:
\begin{equation}
    \label{Overlap}
   |\bra{\theta_{1}}\ket{\theta_{2}}|^{2}=\cos\left(N\Delta\theta \right)^{2}=\mathcal{F}(\ket{\theta_{1}},\ket{\theta_{2}}),
\end{equation}
where $\Delta\theta=\theta_{1}-\theta_{2}$ and $\hbar=1$. We notice that Eqn.~\eqref{Overlap} approaches 1 for any $\Delta \theta = \frac{2\pi k}{N}$, where $k$ is some integer. This will cause the loss function to oscillate or cause $\theta_{2}$ to converge to the wrong value instead of $\theta_{1}$. Thus, for our protocol to converge to the correct value, $|\Delta \theta| \leq \frac{\pi}{2N}$. As $N$ gets larger, this requires an increasingly accurate guess of $\theta_{2}$. We can accomplish this by using a cascade approach: we begin with a random guess for $\theta_{2}$ and use a small $N$ GHZ state to converge to its final value. We then use this final value of the small GHZ state as the starting guess for $\theta_{2}$ for a slightly larger GHZ state. We continue this pattern for steadily larger-$N$ qubit GHZ states for $k$ rounds until we reach a point where $N$ is large enough to yield diminishing results. This can be seen by looking for a vanishing gradient. We then return the $(k-1)^{th}$ value for $\theta_{2}$ as our estimate for $\theta_{1}$. This approach is summarized in Fig~\ref{cascade}.

\section{Comparison to Traditional Metrology Protocol}
\label{sec:tradition}

We compare our approach with a typical metrology approach for GHZ states, as outlined in \cite{11396347}. In this protocol, the GHZ state is prepared and allowed to evolve for a total time $t$, which is split into small intervals. At each time interval, the GHZ state's $\hat{X}$-stabilizer is measured, and the probability of returning a $+1$ outcome is calculated. This probability oscillates with frequency relative to $\theta$, where $\theta=\frac{f \pi}{N}$ and $f$ is the frequency. The frequency is extracted by taking the FFT of the probability function and retaining the peak with the largest amplitude. For this comparison, we use an $N=3$ GHZ state, $\theta=0.23$, a dephasing noise rate of $\gamma=0.11$, and 2500 shots per time step with  200 time steps to simulate the protocol. We also simulate at unit time or when $t=1$. We use QuTiP's \texttt{mesolve} \cite{qutip5} function to evolve the probe state and simulate the measurement protocol, which we then use to extract the probability distribution used to find $\hat{\theta}$. 

\begin{figure}[!t]
    \centering
    \resizebox{\linewidth}{!}{
    \input{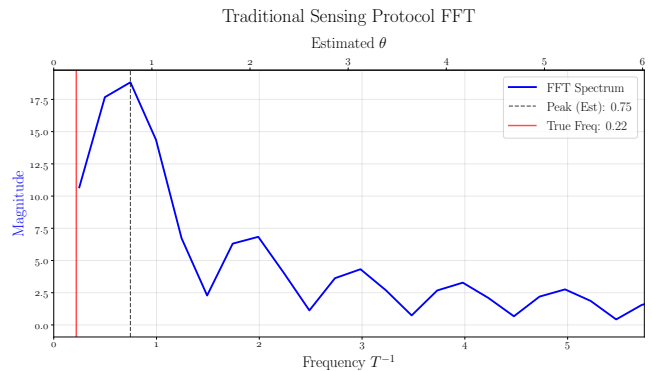}
    }
    \caption{FFT for extracting $\hat{\theta}$ from traditional protocol. For the $N=3$ probe with noise rate $\gamma=0.11$, $\theta=0.23$, 2500 shots per time step, and 200 timesteps for a total time of $t=1$. We find $\hat{\theta}=0.75$, which is within the SQL for an $N=3$ GHZ state. In comparison, our protocol, using the same parameters, yields $\hat{\theta}_{VISTA}=0.223$.}   
    \label{fig:fftdiagram}
\end{figure}

The results, as given in Fig.~\ref{fig:fftdiagram}, show that the traditional protocol's precision in $\theta$ is slightly better than the SQL-based Cram\'er-Rao bound, with the Cram\'er-Rao bound in this case being $\frac{1}{\sqrt{3}}$, versus the actual absolute error between the extracted $\hat{\theta}$ and the actual $\theta$ being approximately 0.55. In contrast, our VISTA protocol with the same parameters yields $\hat{\theta}_{VISTA}=0.223$, a significant improvement. To match this level of absolute error, the traditional protocol can either be evolved for significantly more time, which will lead to decoherence becoming a problem, or more qubits can be used. This shows that our protocol can achieve low absolute-error measurements while being significantly more resource-efficient than traditional methods. As mentioned in Section~\ref{sec:protocoloverview}, the intuition for this advantage is that each measurement for the traditional protocol at each time step is independent of any other time step. This means that the information accumulated per time step is limited to the achievable uncertainty via the Cram\'er-Rao bound per shot. When using an optimizer like ADAM, each epoch depends on information gathered during the previous epoch, which accumulates as the optimizer converges. This allows the optimizer to have far less error in $\theta$ than the traditional approach, as long as the gradients are stable.

\section{Noisy Circuit Ansätz}
\label{sec:Noisycircansätz}

In this section, we discuss using an ansätz that accounts for the channel as well, instead of just using a pure GHZ state. This allows us to learn the noise rate of the probe state, giving us a way to learn the entire state, not just the signal parameters. To do this, we sharpen the gradients of our loss function via a technique we call \textit{Quasi-Normalization}, discussed in Section~\ref{sec:performance noisy}. We will focus on two channels relevant to quantum magnetometry, mainly the dephasing and amplitude damping channels, and introduce the circuits and conditions for those circuits to match the Lindbladian evolution of each case.

\subsection{Dephasing Case}

Given the following probe state $\rho=\ket{\mathrm{GHZ}}\bra{\mathrm{GHZ}}$, where $\ket{\mathrm{GHZ}}=\frac{1}{\sqrt{2}}\left(\ket{0}^{\otimes N}+\ket{1}^{\otimes N}\right)$, we subject the probe to the following evolution:
\begin{equation}
    \label{Lindbladian}
    \frac{\partial \rho}{\partial t}=\mathcal{L}(\rho)
    \coloneqq -\frac{i}{\hbar}[\hat{H},\rho]\theta+\gamma \left(\sum_{j=1}^{N} \hat{Z}_{j}\rho \hat{Z}_{j} -\rho \right),
\end{equation}
where $\gamma$ is the noise rate, and the Hamiltonian is given by \eqref{hamiltonian}.
Setting $\hbar=1$, and solving \eqref{Lindbladian} for unit time ($t=1$), we see that:
\begin{align}\label{ls}
\rho(\theta,\gamma) &= \frac{1}{2}\Big( \ket{0}^{\otimes N}\bra{0}^{\otimes N}  \nonumber \\
 &+ e^{-2N(i\theta +\gamma)} \ket{0}^{\otimes N}\bra{1}^{\otimes N} \nonumber \\
 &+ e^{-2N(-i\theta +\gamma)} \ket{1}^{\otimes N}\bra{0}^{\otimes N}   
+ \ket{1}^{\otimes N}\bra{1}^{\otimes N} \Big).
\end{align}

\begin{figure}[!t] 
    \centering
    \includegraphics[width=0.5\textwidth, height=7cm, keepaspectratio]{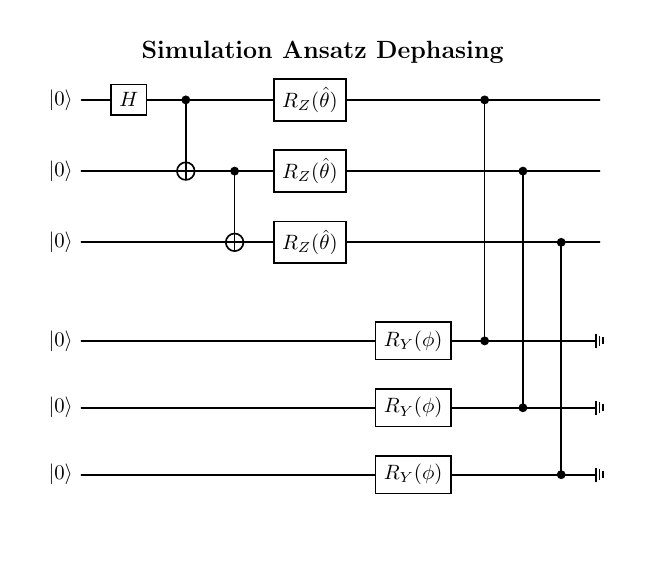} 
    \caption{Circuit for the $N=3$ Dephasing ansätz.}
    \label{circuitnoisydephasing}
\end{figure}

Since we notice that the noise and signal terms commute for the Lindbladian super-operator, we can write the evolution as:
\begin{equation}
    \label{commutedephase}
    \rho(\theta,\gamma)=e^{\mathcal{L}(\rho)}\rho=e^{\mathcal{L}_{sig}(\rho)}e^{\mathcal{L}_{channel}(\rho)}\rho,
\end{equation}
which means we can use the circuit ansätz given in Fig.~\ref{circuitnoisydephasing}, shown for the $N=3$ case.
After tracing out the ancilla qubits, the output state of the circuit is given by:
\begin{align}
\label{circuitoutput}
\rho_{circ}(\hat{\theta},\phi) &= \frac{1}{2}\Big( \ket{0}^{\otimes N}\bra{0}^{\otimes N}  \nonumber \\
 &+ e^{-2Ni\hat{\theta}}\cos^{N}\phi \ket{0}^{\otimes N}\bra{1}^{\otimes N} \nonumber \\
 &+ e^{2Ni\hat{\theta}}\cos^{N}\phi \ket{1}^{\otimes N}\bra{0}^{\otimes N}  \nonumber \\ 
 &+ \ket{1}^{\otimes N}\bra{1}^{\otimes N} \Big).
\end{align}
Here, we see that \eqref{ls} and \eqref{circuitoutput} are equal if:
\begin{align}\label{equalitycondition}
    \theta &= \hat{\theta}, \nonumber \\
    \cos{\phi} &=e^{-2\gamma}; \ 0 \leq \phi < \frac{\pi}{2}.
\end{align}

\subsection{Amplitude Damping Channel}

Following the same argument as the previous case, we define the Linbladian for the Amplitude Damping channel as:
\begin{equation}
    \label{linbladampli}
\frac{\partial \rho}{\partial t}= -\frac{i}{\hbar}[\hat{H},\rho]\theta+\gamma \left(\sum_{j=1}^{N} \sigma_{-j}\rho \sigma^{\dagger}_{-j} -\frac{1}{2}\{ \sigma_{-j}\sigma^{\dagger}_{-j},\rho \}\right),
\end{equation}
where $\sigma_{-}=\ket{0}\bra{1}$. 
To solve this, we notice that the channel acts on each qubit independently. Solving the single qubit case for a $\ket{+}$ state yields \cite{manzano2020short}:
\begin{align}
    \label{singlequbitampl}
    \rho_{+}&(\theta,\gamma)=\left(1-\frac{1}{2}e^{-\gamma}\right)\ket{0}\bra{0}\nonumber \\&+\frac{1}{2}e^{-\frac{\gamma}{2}}\left(e^{-2i\theta}\ket{0}\bra{1}+e^{2i\theta}\ket{1}\bra{0}\right)+\frac{1}{2}e^{-\gamma}\ket{1}\bra{1}.
\end{align}
We can now use the action of the channel on each term of \eqref{singlequbitampl} to see that:
\begin{equation}
\label{exactampdampout}
\begin{split}
\rho(\theta,\gamma) =& \frac{1}{2} \ket{0}^{\otimes N}\bra{0}^{\otimes N} + \frac{1}{2} e^{-\frac{N\gamma}{2}} e^{-2iN\theta} \ket{0}^{\otimes N}\bra{1}^{\otimes N} \\
&+ \frac{1}{2} e^{-\frac{N\gamma}{2}} e^{2iN\theta} \ket{1}^{\otimes N}\bra{0}^{\otimes N} \\
&+ \frac{1}{2} \sum_{x \in \{0,1\}^N} (e^{-\gamma})^{|x|} (1-e^{-\gamma})^{N-|x|} \ket{x}\bra{x}.
\end{split}
\end{equation}

\begin{figure}[!t] 
    \centering
    \includegraphics[width=0.5\textwidth]{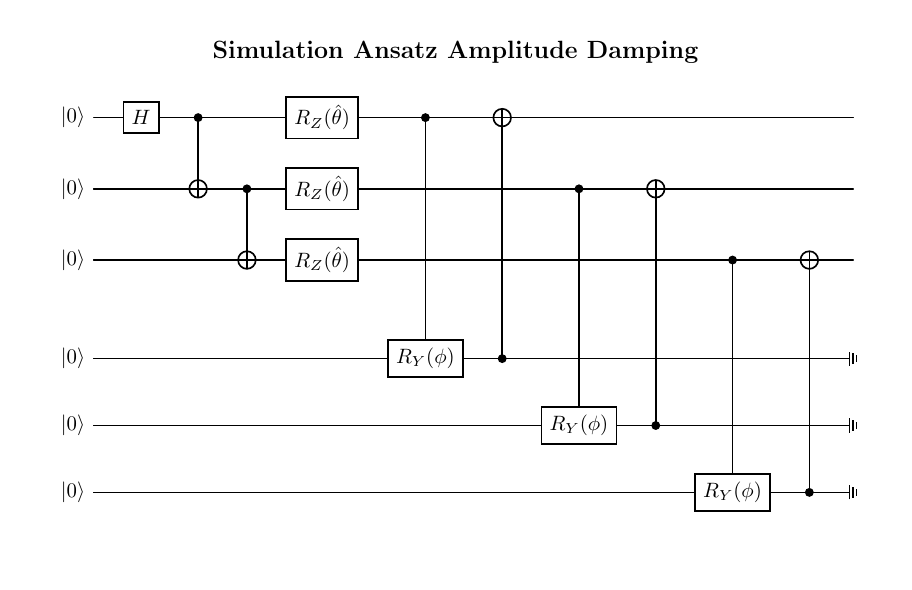} 
    \caption{Circuit for the $N=3$ Amplitude Damping ansätz.}
    \label{circuitnoisyampli}
\end{figure}

Since the signal and noise terms of the Lindblad super-operator commute, we can again use a circuit ansätz that follows the form of \eqref{commutedephase} as seen in Fig.~\ref{circuitnoisyampli} for the $N=3$ case.
We now write the output of this circuit as:
\begin{equation}
\label{eq:ampdampcircout}
\begin{split}
    \rho(\hat{\theta},\phi)_{\text{circ}} &= \frac{1}{2} \Big( \ket{0}^{\otimes N}\bra{0}^{\otimes N} \\
    &\quad + \sum_{x \in \{0,1\}^N} \alpha^{N-|x|} (1-\alpha)^{|x|} \ket{x}\bra{x} \Big) \\
    &+ \frac{1}{2} (1-\alpha)^{N/2} \Big( e^{-i 2N\theta} \ket{0}^{\otimes N}\bra{1}^{\otimes N} \\
    &\quad + e^{i 2N\theta} \ket{1}^{\otimes N}\bra{0}^{\otimes N} \Big),
\end{split}
\end{equation}
where $\alpha=\sin^{2}(\phi)$.
We get an exact match to \eqref{exactampdampout} if $\cos(\phi)=e^{-\frac{\gamma}{2}}$.
\begin{remark}
\label{remark1}
For general channels, the condition \eqref{commutedephase} is not true, which means the circuit must be Trotterized for an accurate estimate. For $d$-rounds of Trotter-steps, this would require $O(Nd)$ ancilla qubits, which may not be ideal. However, in general the type of circuit used for the dephasing and amplitude damping case can be seen as unitary evolution of the probe state followed by a discrete application of the channel. For sufficiently small times, there is little difference between this sequential circuit and the continuous Lindblad style evolution, meaning that one can only use $O(N)$ ancillas to get a rough approximation of the noise rate. This can be improved by training a small neural-network to correct the learned values of the noise rate $\gamma$ to more accurate values. 
\end{remark}

\section{Performance Comparison between Pure and Noisy Circuit Ansätze for learning $\theta$ and $\gamma$}
\label{sec:performance noisy}

Now, we turn to the question of when learning the channel noise parameters is possible without losing precision for $\theta$. We will consider two cases: dephasing as before, and an amplitude damping channel, with the key difference between the two being that although the dephasing channel dampens the amplitude of the coherence terms of the GHZ state, the state still has a form similar to a GHZ state, while the amplitude damping channel adds extra terms not originally found in the GHZ state. As we will see, the presence of these extra terms will determine whether learning the noise helps or hurts the protocol in finding $\theta$ near the minima of the loss function. We note that we focus on single-shot performance here, i.e., $\sqrt{\nu}=1$.

\subsection{Dephasing Case}

Since we use the swap test to calculate the Hilbert-Schmidt inner product $\mathrm{Tr}\left(\rho(\theta,\gamma)\rho_{circ}(\theta,\phi)\right)$ as a proxy for the fidelity between the states, we calculate $\mathcal{Q}_{HS}(\rho_{\theta})$, which is an approximation of the QFI for our protocol, where  $2\mathcal{Q}_{HS}(\rho_{\theta})\leq\mathcal{Q}(\rho_{\theta})$ and $\mathcal{Q}(\rho_{\theta})$ is the typical QFI given by the Uhlmann Fidelity \cite{paris2009quantum}. We derive this bound in Appendix~\ref{sec:AppenA}.

\subsubsection{Unnormalized Case}

We first look at the unnormalized trace overlap to justify our normalization scheme. Using the conditions in \eqref{equalitycondition}, where $\phi$ is matched to correspond to some $\gamma'$, we see that
\begin{align}
    \label{traceoveralap}
    &\mathrm{Tr}\left(\rho(\theta_{1},\gamma)\rho_{circ}(\theta_{2},\phi_{\gamma'})\right)= \nonumber \\ &\frac{1}{2}\left(1+e^{-2N(\gamma+\gamma')}\cos(2N\Delta\theta)\right),
\end{align}
where $\Delta\theta=\theta_{1}-\theta_{2}$. We now see that:
\begin{align}
    \label{approxdephaseqfi}
   \mathcal{Q}_{HS}(\rho_{\theta})&\coloneqq \lim_{\Delta \theta \rightarrow 0}  -\frac{\partial^2 \mathrm{Tr}\left(\rho(\theta_{1},\gamma)\rho_{circ}(\theta_{2},\phi_{\gamma'})\right) }{\partial \Delta \theta^2} \nonumber \\ \nonumber \\  &= 2e^{-2N(\gamma+\gamma')}N^2
\end{align}
Inserting into the Cram\'er-Rao bound $\delta\theta \geq \frac{1}{\sqrt{2 \mathcal{Q}_{HS}(\rho_{\theta})}}$, for $\gamma'=\gamma$ we see that:
\begin{equation}
\label{crunnorm}
    \delta \theta \geq \frac{e^{2N\gamma}}{2N},
\end{equation}
 where in general for the swap test with multiple shots, there will be an extra factor of $\frac{1}{\sqrt{\nu}}$, where $\nu$ is the number of shots. In the unnormalized case, we see that learning $\phi_{\gamma}$ for the dephasing channel significantly worsens our ability to learn $\theta$ with high precision near the minima. In the case of dephasing, we maximize our ability to find $\theta$ with high precision by assuming a pure ansätz. But, as we will see later, if we are willing to sacrifice some precision of $\theta$, this means we can learn both signal and noise simultaneously, thus learning the entire state, which is enhanced by Quasi-Normalization, as we shall see next.

\subsubsection{Quasi-Normalized Case}

\begin{figure}[t]
\centering
\resizebox{\linewidth}{!}{
    \input{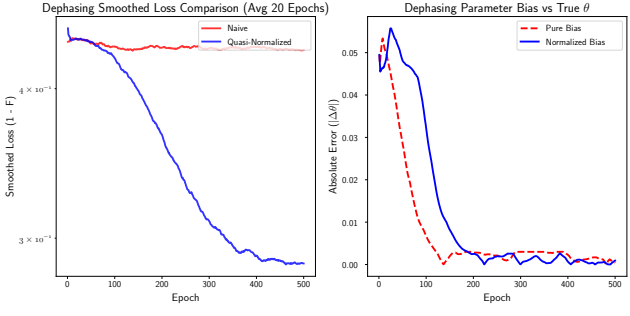}
}
\caption{Comparing the pure ansätz to the Quasi-Normalized ansätz for dephasing for $\gamma=0.1$, $\theta=10^{-3}$, $N=10$ qubits. We see that the gradients of the Quasi-Normalized loss function are much sharper, leading to better convergence. Also, although the pure case may perform better near the minimum, we see little difference in error between the two approaches.}
\label{fig:dephasingbias}
\end{figure}

Unlike the Uhlmann Fidelity, the Hilbert Schmidt Inner Product is bounded by the purity $\mathrm{Tr}(\rho^2)$ of the state in question. We can normalize the Hilbert Schmidt Inner Product $\mathrm{Tr}\left(\rho(\theta_{1},\gamma)\rho_{circ}(\theta_{2},\phi_{\gamma'})\right)$ by:
{\small
\begin{equation}
    \label{normalizationdephasing}
    \mathrm{Tr}\left(\rho(\theta_{1},\gamma)\rho_{circ}(\theta_{2},\phi_{\gamma'})\right) \rightarrow \frac{\mathrm{Tr}\left(\rho(\theta_{1},\gamma)\rho_{circ}(\theta_{2},\phi_{\gamma'})\right)}{\sqrt{\mathrm{Tr}(\rho(\theta_{1},\gamma)^2)\mathrm{Tr}(\rho_{circ}(\theta_{2},\phi_{\gamma'})^2)}}.
\end{equation}
}
However, we do not know $\rho(\theta_{1},\gamma)$. It is sufficient to set the purity of the unknown state to some constant, in this case $1$, and continue with a Quasi-Normalized version of the Hilbert Schmidt Inner Product:
{\small
\begin{equation}
\label{normoverlap}
    \mathrm{Tr}\left(\rho(\theta_{1},\gamma)\rho_{circ}(\theta_{2},\phi_{\gamma'})\right)_{QN}=\frac{1+e^{-2N(\gamma+\gamma')}\cos(2N\Delta\theta)}{\sqrt{2(1+e^{-4N\gamma'})}},
\end{equation}
}
where we now see that inserting \eqref{normoverlap} into \eqref{approxdephaseqfi} with $\gamma'=\gamma$ leads to:
\begin{equation}
   \mathcal{Q}_{HSQN}(\rho_{\theta}) =\frac{4e^{-4N\gamma}N^2}{\sqrt{2(1+e^{-4N\gamma})}}.
\end{equation}
Then, inserting this into the Cram\'er-Rao bound yields:
\begin{equation}
\label{crboundnormed}
    \delta \theta_{QN}\geq \frac{e^{4N\gamma}}{2N} \left( \frac{1 + e^{-4N\gamma}}{8} \right)^{\frac{1}{4}},
\end{equation}
where we see a significantly better scaling bound than the unnormalized case. However, it is still not as good as using a pure state ansätz for dephasing near the minimum. But, as shown in Fig.~\ref{fig:dephasingbias}, the sharper gradients from Quasi-Normalization make this difference minimal in practice.

\begin{figure}[t]
\centering
\resizebox{\linewidth}{!}{
    \input{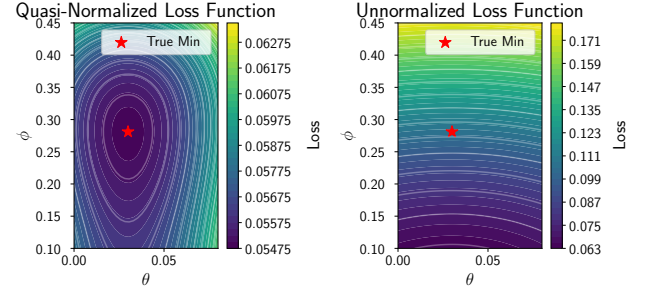}
}
\caption{Comparison between the Unnormalized and Quasi-Normalized loss functions. We notice that the minima have been biased towards the correct noise rate compared to the unnormalized case, allowing us to find a noise value closer to the actual case.  We also note that the Quasi-Normalized loss function has significantly sharper $\theta$ gradients than the Unnormalized case, leading to better convergence behavior.}
\label{fig:losplots}
\end{figure}

By normalizing the trace overlap, we force the optimizer to learn the noise to maximize the overlap, biasing it towards the actual state, as shown in Fig.~\ref{fig:losplots}. It is important to remember that this performance comparison measures the sharpness of the curvature near the minimum for the different loss functions and does not necessarily guarantee sharper gradients farther away. As we see from the previous Fig.~\ref{fig:losplots}, the minimum may be shallower than the pure case, but the Quasi-Normalization sharpens the gradients significantly.

\subsection{Amplitude Damping Case}

\begin{figure}[!t]
\centering
\resizebox{\linewidth}{!}{
    \input{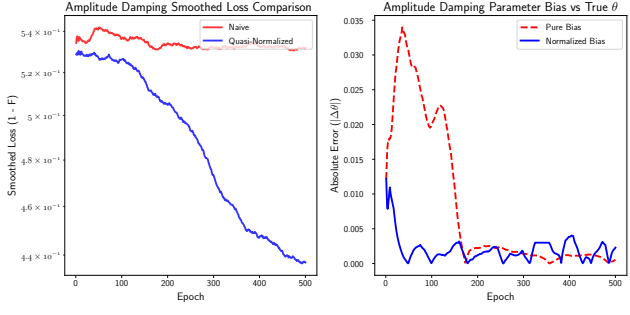}
}
\caption{Comparing the pure ansätz to the Quasi-Normalized ansätz for amplitude damping for $\gamma=0.2$, $\theta=10^{-3}$ ,$N=10$ qubits. We see that the gradients of the Quasi-Normalized loss function are much sharper than the pure case, albeit less so than the dephasing case, leading to better convergence. We see that the pure case and the Quasi-Normalized case have comparable error at the end of the epochs, meaning we can learn $\gamma$ without sacrificing $\theta$'s absolute error.}
\label{fig:ampddampbias}
\end{figure}

To compare the performance between the pure ansätz and the Quasi-Normalized noisy ansätz, we first look at the Hilbert-Schmidt inner product between the pure ansätz and the target state, following the approach of the previous section. 
\begin{equation}
\label{overlappuraamp}
\begin{split}
\mathrm{Tr}(\rho_{target}(\theta_{1},\gamma) \rho_{pure}(\theta_{2})) =& \frac{1}{4} + \frac{1}{4}(1 - e^{-\gamma })^N + \frac{1}{4}e^{-N\gamma} \\
&+ \frac{1}{2}e^{-\frac{N\gamma}{2}} \cos(2N(\theta_{1} - \theta_{2})).
\end{split}
\end{equation}
Inserting into \eqref{approxdephaseqfi}, we see that:
\begin{equation}
    \label{qfamppure}
\mathcal{Q}_{HSpure}(\rho_{\theta})=2e^{-\frac{N\gamma}{2}}N^2.
\end{equation}
We now compare this to using the output of the circuit ansätz \eqref{eq:ampdampcircout} in \eqref{normalizationdephasing} with $\mathrm{Tr}(\rho(\theta_{1},\gamma)^2)=1$. 
\begin{align}
\label{quasinormampdamp}
\mathrm{Tr}&\left(\rho(\theta_{1},\gamma)\rho_{\mathrm{circ}}(\theta_{2},\phi_{\gamma'})\right)_{QN} = \frac{\mathcal{N}}{\sqrt{\mathcal{D}}}, \nonumber \\
\mathcal{N} &= \frac{1}{4} + \frac{1}{4}(1-e^{-\gamma })^N + \frac{1}{4}(1-e^{-\gamma' })^N \nonumber \\
&\quad + \frac{1}{4}\big(e^{-(\gamma+\gamma')} + (1-e^{-\gamma })(1-e^{-\gamma' })\big)^N \nonumber \\
&\quad + \frac{1}{2} e^{-\frac{N(\gamma+\gamma')}{2} } \cos(2N(\theta_1 - \theta_2)),\nonumber \\ \nonumber\\ 
\mathcal{D} &= \frac{1}{4} + \frac{1}{2}(1 - e^{-\gamma' })^N + \frac{1}{2}e^{-N\gamma' } \nonumber \\
&\quad + \frac{1}{4}\big(e^{-2\gamma' } + (1 - e^{-\gamma' })^2\big)^N.
\end{align}
Setting $\gamma=\gamma'$, and inserting into \eqref{approxdephaseqfi}, we see that:
\begin{equation}
    \mathcal{Q}_{HSQN}(\rho_{\theta})=\frac{2e^{-N\gamma} N^2}{\sqrt{D}}.
\end{equation}
At first glance, it may look like the curvature of the  minima for the pure case is again steeper than the Quasi-Normalized case, but if we take the ratio $\frac{\mathcal{Q}_{HSQN}(\rho_{\theta})}{\mathcal{Q}_{HSpure}(\rho_{\theta})}$, and expand for small $\gamma$, we see that:
\begin{equation}
    \label{qfiratio}
    \frac{\mathcal{Q}_{HSQN}(\rho_{\theta})}{\mathcal{Q}_{HSpure}(\rho_{\theta})}=\frac{e^{-\frac{N\gamma}{2}}}{\sqrt{D}} \approx 1-\frac{N(N+1)}{8}(\gamma)^2+O(\gamma^3),
\end{equation}
which means the Quasi-Normalization helps recover most of the curvature sharpness lost from the noise channel. As we see in Fig.~\ref{fig:ampddampbias}, our convergence behavior is much improved with the Quasi-Normalization approach, and we have a comparable absolute error in $\theta$ compared to the pure ansätz. This is beneficial, as we can learn $\theta$ and $\gamma$ simultaneously without too much error in learning $\theta$.

\begin{remark}
For more general channels where deriving the purity may be infeasible, it is unnecessary to use an analytical version of the purity for Quasi-Normalization to work. Since the Hamiltonian does not increase the entanglement of the GHZ state, even for the multi-parameter generalization, and the GHZ state has a low, constant Schmidt-Rank for any $N$, the purity can be efficiently calculated exactly by a tensor-network, and applied to the cost function after calculating the Hilbert-Schmidt inner product via the swap test. In general for an MPS tensor-network, the time complexity for an inner-product between two $N$ site MPS networks as in the case for states represented in the doubled Hilbert space is $O(4N\chi^3)$, where $\chi$ is the bond dimension of the network. Since the entanglement does not grow during this evolution, and $\chi=2$ for an $N$-qubit GHZ state in the doubled representation, calculating the purity becomes a linear-time calculation. 
\end{remark}

\subsection{Noisy Ansätz Performance for Learning Dephasing Noise Rate \texorpdfstring{$\gamma$}{gamma}}

\begin{figure}[!t] 
    \centering
    \includegraphics[width=0.5\textwidth, height=6cm, keepaspectratio]{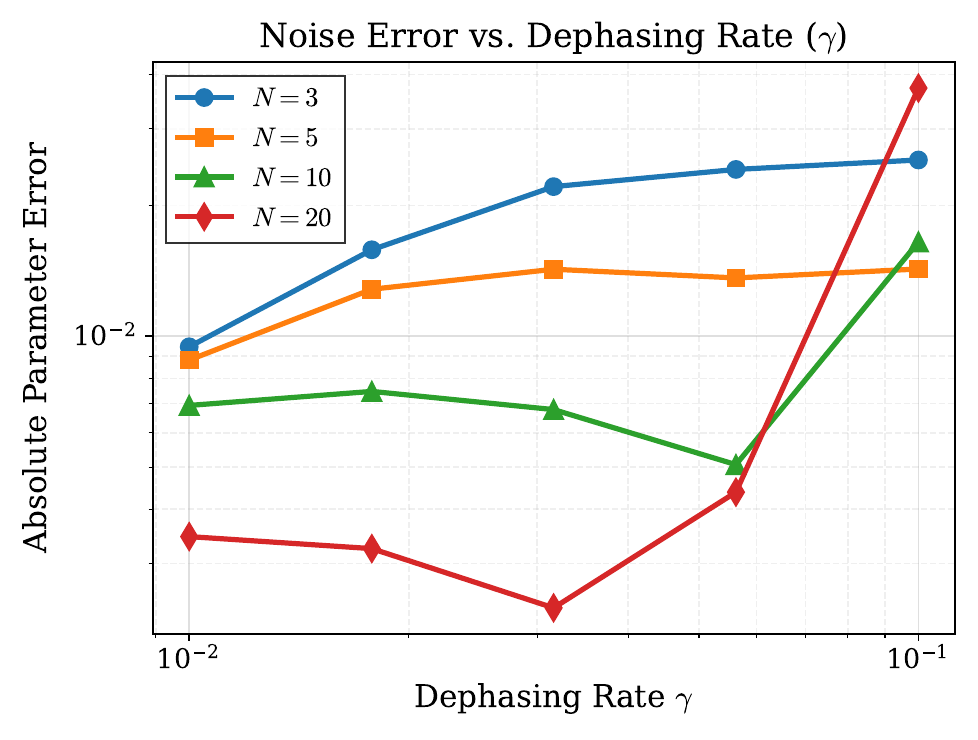} 
    \caption{Averaged absolute error for a typical range of $\gamma$ values for different-sized GHZ states with $\theta=0.05$. We use a shot scheduler ranging from $10$K to $40$K qubits per epoch.  We notice that due to the higher noise rate, there are diminishing performance returns past $N=10$, with an $N=20$ GHZ state performing worse than an $N=3$ GHZ state nearing $\gamma=0.1$. }
    \label{abserrorgamma}
\end{figure}

As with Fig.~\ref{abserrortheta}, we notice in Fig.~\ref{abserrorgamma} that the absolute error in $\gamma$ decreases with larger $N$ until the point of diminishing returns, typically when $\gamma$ is large. Here we see that past $\gamma=0.05$, we begin to have lesser performance from $N=10$ GHZ states onward, which is expected due to the form of \eqref{normalizationdephasing}. Better performance can be achieved by deriving a calibration curve between actual and learned $\gamma$ values, or by training a small neural network to post-process the learned values. Additionally, one can use more shots to further reduce the error.

\subsection{Noisy Ansätz Performance for Learning Amplitude 
Damping Noise Rate \texorpdfstring{$\gamma$}{gamma}}

\begin{figure}[!t] 
    \centering
    \includegraphics[width=0.5\textwidth, height=6cm, keepaspectratio]{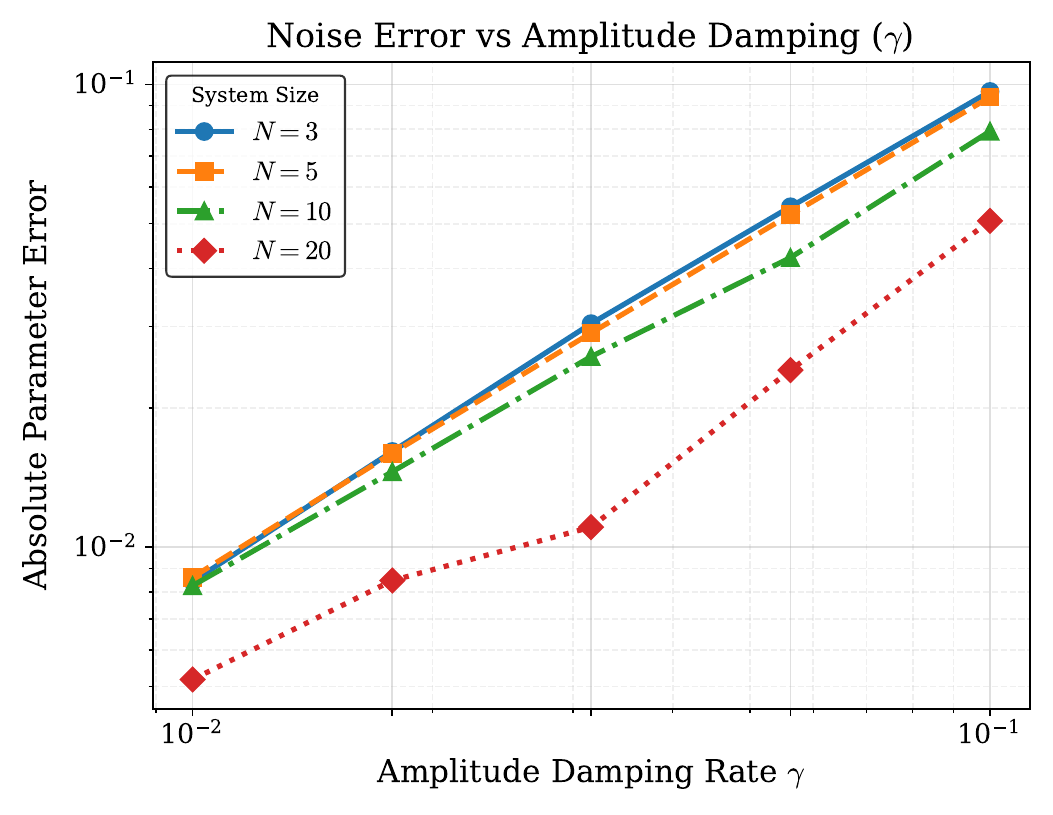} 
    \caption{Averaged absolute error for a typical range of $\gamma$ values for different-sized GHZ states with $\theta=0.05$. We use a shot scheduler ranging from $10$K to $40$K qubits per epoch. Unlike the dephasing case, we notice no diminishing returns as $N$ approaches 20, but the relative error is higher due to the shot count. Since the $\gamma$ error curves are monotonically increasing, it is simple to account for this via a calibration curve. }
    \label{abserrorgammaampdamp}
\end{figure}

Unlike the dephasing case, in Fig.~\ref{abserrorgammaampdamp} we notice no diminishing returns over the range of qubit sizes as $N$ approaches 20 qubits, though the relative error is high. As in the dephasing case, since the $\gamma$ error curves are monotonically increasing, one can use a calibration procedure to accurately estimate the true $\gamma$ from the protocol's estimate; however, a larger number of shots can also reduce this error.

\section{Multi-Parameter Vector Metrology}
\label{sec:multiparam}

We now generalize the VISTA protocol to vector metrology. We now consider the following Lindbladian:
\begin{align}
    \label{Linbladianmulti}
    \frac{\partial \rho}{\partial t}=-\frac{i}{\hbar}  [\vec{H} \cdot \vec{\theta},\rho]+\sum_{k}\left(L_{k}\rho L_{k}^{ \dagger}-\frac{1}{2}\{L_{k}^{\dagger}L_{k},\rho\}\right),
\end{align}
where we consider the two-parameter case such that $\vec{H}=\left[\sum_{j}\hat{Z}_{j},\sum_{j}\hat{X}_{j}\right]$, $\vec{\theta}=\left[\theta_{1},\theta_{2}\right]$. As shown in Section~\ref{sec:tradition}, the optimal measurement for a $\hat{Z}$-Hamiltonian is the GHZ state's $\hat{X}$-stabilizer. For the $\hat{X}$-Hamiltonian, we can derive the optimal measurement by first deriving the GHZ state's QFI with respect to it. This can be given as \cite{Ouyang-pra23}:
\begin{equation}
    \label{qfi bound purestate}
    \mathcal{Q}\left(\rho_\theta,\frac{\partial \rho_\theta}{\partial \theta}\right) \leq 4\mathrm{Tr}\left(\rho \hat{H}^{2}\right) - 4\left(\mathrm{Tr}\left(\rho \hat{H}\right)\right)^{2},
\end{equation}
where inserting the GHZ state and $\hat{X}$ yields:
\begin{equation}
    \label{xhamiltonianqfi}
    4\mathrm{Tr}\left(\rho \hat{H}^{2}\right) - 4\left(\mathrm{Tr}\left(\rho \hat{H}\right)\right)^{2}=4N.
\end{equation}
Thus, the GHZ state has SQL level sensitivity to the $\hat{X}$-Hamiltonian. Since the state's entanglement does not add to its $\hat{X}$-Hamiltonian sensitivity, the optimal measurement for this case is $\bigotimes_{j}\hat{Z}_{j}$. However, we now see an issue: for $N$ even, $\left[\bigotimes_{j}\hat{X}_{j},\bigotimes_{j}\hat{Z}_{j}\right]=0$; but for $N$ odd, $\{\bigotimes_{j}\hat{X}_{j},\bigotimes_{j}\hat{Z}_{j}\}=0$. This means that given these measurement operators, simultaneous optimal measurement of both $\theta_{1}$ and $\theta_{2}$ is only possible for even-$N$ GHZ states. 

\begin{figure}[!t] 
    \centering
    \includegraphics[width=0.5\textwidth]{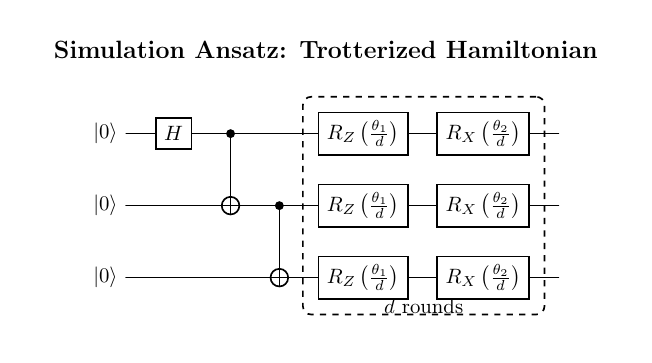} 
    \caption{Circuit for the $N=3$ Trotter Vector Metrology ansätz.}
    \label{trotter}
\end{figure}

\begin{figure}[!t] 
    \centering
    \includegraphics[width=0.5\textwidth]{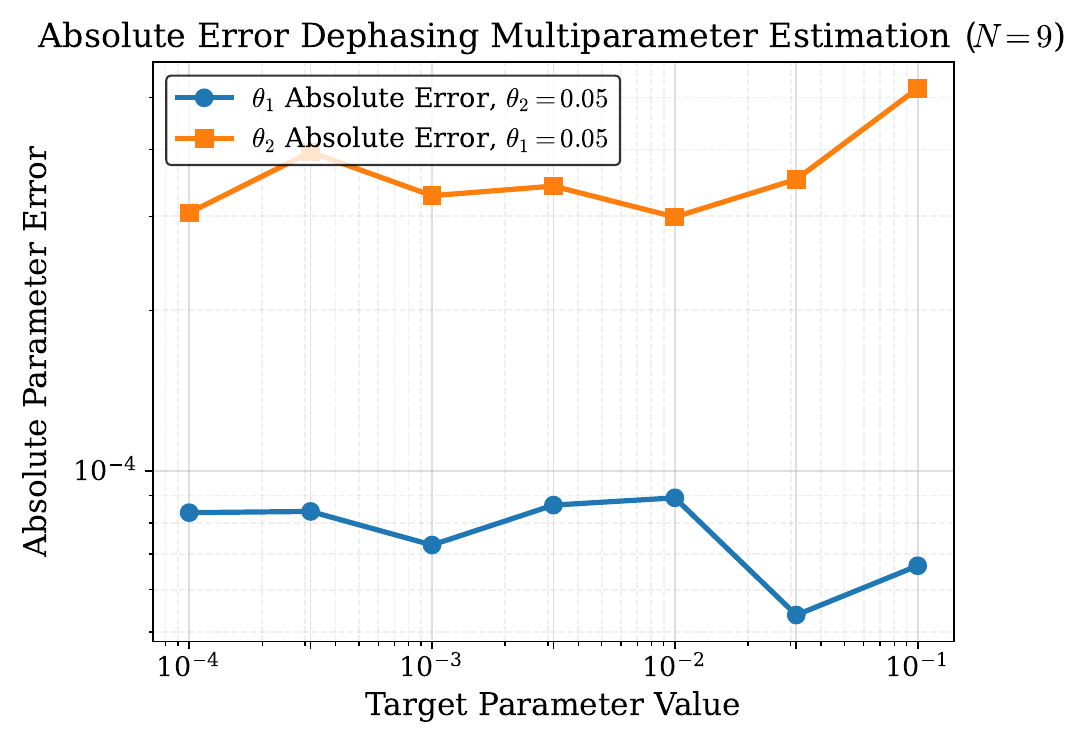} 
    \caption{Averaged absolute error for $\theta_{1}$, $\theta_{2}$, respectively with the other held constant for a $N=9$ qubit GHZ state with dephasing rate $\gamma=0.02$. This plot shows significantly lower absolute error for $\theta_{1}$ than for $\theta_{2}$ due to GHZ states being more sensitive to the $\hat{Z}$ part of the Hamiltonian.}
    \label{abserrormultiparam}
\end{figure}

Since VISTA uses the swap test, a non-local entangling measure, we can circumvent this restriction on the set of local measurement operators. Although the unitary generated by $\vec{H}$ is transversal for each qubit, the single qubit unitaries cannot be written as a product since $\{\hat{X}_{j},\hat{Z}_{j}\}=0$. Therefore, we use a Trotter-circuit ansätz for the pure case as shown in Fig.~\ref{trotter}. We now demonstrate VISTA's ability to extract both $\theta_{1}$ and $\theta_{2}$ simultaneously for $N=9$ qubit state in Fig.~\ref{abserrormultiparam}, where we again notice low absolute error for both parameters. We also notice that the absolute error for $\theta_{1}$ is significantly smaller than for $\theta_{2}$, which stems from the GHZ state's higher sensitivity for the $\hat{Z}$ part of the Hamiltonian. It is also possible to use a circuit ansätz for the vector metrology case that learns the noise rate via the Quasi-Normalization technique described in Section~\ref{sec:performance noisy}, but since the Hamiltonian does not commute with the dephasing channel in this case, a more accurate estimate of $\gamma$ will require a Trotter-approximation of the channel, as mentioned in Remark~\ref{remark1}.

\section{Discussion}
\label{sec:discussion}

In this work, we introduce the VISTA protocol for extracting the signal and noise parameters from a passively evolving GHZ state. We demonstrate the protocol's exceptional ability to estimate a parameter $\theta$ with low absolute error. By leveraging variational quantum circuits and swap-test measurements, we enable gradient-based optimization to reliably converge on the true parameter value. We also demonstrated that, with a moderate noise rate $\gamma$, at the expense of a high number of shots, we can achieve near-Heisenberg error scaling in $N$, for a finite range of values. However, asymptotically for large $N$, it will approach the SQL due to noise. We also demonstrate how Quasi-Normalization allows us to use low-parameter-count circuit ansätze to learn both $\theta$ and $\gamma$ simultaneously, without vanishing-gradient issues and with minimal error penalty on $\theta$. Finally, we briefly discuss how to generalize VISTA to multi-parameter vector metrology. 

We believe this work represents an exciting and practical approach to quantum metrology due to its relative simplicity and its use of a quantum computer in a practical application. The main bottleneck of the approach is the swap test itself, but there have been a number of new approaches aiming to replace the swap test with a protocol able to compute measures similar to the Hilbert-Schmidt inner product but with fewer samples \cite{wiebe2014quantum,PhysRevLett.124.060503,fang_et_al:LIPIcs.ESA.2025.4}. Specifically, \cite{wiebe2014quantum,fang_et_al:LIPIcs.ESA.2025.4} claim sampling complexity that scales as $O(\frac{1}{\epsilon})$ time instead of  $O(\frac{1}{\epsilon^2})$ time. This would allow \ref{crupure} to scale with $\frac{1}{\nu}$, enabling similar performance with significantly fewer shots. It remains to be seen whether these protocols would be feasible in practice, given their added circuit complexity. Future work may lead us to gauge VISTA's performance in the Early-Fault-Tolerant regime, where the channel governing the quantum computer's evolution may degrade performance. Finally, we note that the closed-loop nature of our protocol makes it useful for distributed quantum sensing applications \cite{zhang2021distributed}.

\section*{Acknowledgments}
We wish to thank our QBEAM project collaborators and Amit Ashok for their insights and feedback. The work of O.~N., N.~R., and C.~G. was supported by the U.S. Army Research Office Grant W911NF-24-1-0080.

\appendix

\begin{appendices}

\section{Derivation of QFI Lower bound for Hilbert-Schmidt Inner Product}
\label{sec:AppenA}

We begin with a probe state $\rho$ which we spectrally decompose as $\rho=\sum_{i}p_{i}\ket{i}\bra{i}$.
We then let $\rho \rightarrow e^{\frac{-i}{\hbar}\theta \hat{H}}\rho e^{\frac{i}{\hbar}\theta \hat{H}} \coloneqq \rho_{\theta}$. From here, we can write the typical QFI derived from the Uhlmann Fidelity as \cite{paris2009quantum}:
\begin{equation}
    \label{QFIUHL}
    \mathcal{Q}(\rho_{\theta})=2\sum_{i\neq j}\frac{|\bra{i}\frac{\partial \rho_{\theta}}{\partial \theta}\ket{j}|^{2}}{p_{i}+p_{j}},
\end{equation}
where the diagonal terms are zero. Looking at the Hilbert Schmidt inner product between two states $\mathrm{Tr}\left(\rho_{\theta} \sigma\right)$, we focus on the case of the purity, i.e., when $\sigma=\rho_{\theta}$. The purity of a density operator is invariant under the action of a unitary operator, so we see that:
\begin{equation}
    \label{secondder}
    \frac{\partial^2}{\partial \theta^2}\mathrm{Tr}\left(\rho_{\theta}^2 \right)=0.
\end{equation}
We expand this leading to:
\begin{equation}
    \label{secondder}
    \frac{\partial^2 \mathrm{Tr}\left(\rho_{\theta}^2 \right)}{\partial \theta^2}=2\mathrm{Tr}\left[\left(\frac{\partial \rho_{\theta}}{\partial \theta}\right)^2\right]+2\mathrm{Tr}\left(\rho_{\theta}\frac{\partial^2 \rho_{\theta}}{\partial \theta^2}\right)=0.
\end{equation}
We use \eqref{secondder} to define a new  quantity $\mathcal{Q}_{HS}(\rho_{\theta})$ as:
\begin{align}
    \label{dervativeequality}
    \mathcal{Q}_{HS}(\rho_{\theta})\coloneqq -\lim_{\theta' \rightarrow \theta} \frac{\partial^2}{\partial \theta^2}\mathrm{Tr}\left(\rho_{\theta'}\rho_{\theta} \right)&=-\mathrm{Tr}\left(\rho_{\theta}\frac{\partial^2 \rho_{\theta}}{\partial \theta^2}\right)\\ \nonumber &=\mathrm{Tr}\left[\left(\frac{\partial \rho_{\theta}}{\partial \theta}\right)^2\right].
\end{align}
We now rewrite the trace in \eqref{dervativeequality} over the eigenbases of $\rho$ as:
\begin{equation}
    \label{tracehsfid}
   \frac{\partial^2 \mathrm{Tr}\left(\rho_{\theta}^2 \right)}{\partial \theta^2}=\sum_{i\neq j}|\bra{i}\frac{\partial \rho_{\theta}}{\partial \theta}\ket{j}|^{2}.
\end{equation}
Comparing \eqref{tracehsfid} to \eqref{QFIUHL}, we see that since $p_{i}+p_{j}\leq 1$, $\frac{2}{p_{i}+p_{j}}\geq 2$. Therefore, $\mathcal{Q}(\rho_{\theta})\geq 2  \mathcal{Q}_{HS}(\rho_{\theta})$.

\end{appendices}

\bibliography{cites}

@article{paris2009quantum,
  title={Quantum estimation for quantum technology},
  author={Paris, Matteo GA},
  journal={International Journal of Quantum Information},
  volume={7},
  number={supp01},
  pages={125--137},
  year={2009},
  publisher={World Scientific}
}

@article{qutip5,
  title = {QuTiP 5: The Quantum Toolbox in {Python}},
  author = {
    Lambert, Neill and Gigu{`e}re, Eric and Menczel, Paul and Li, Boxi and
    Hopf, Patrick and Su{'a}rez, Gerardo and Gali, Marc and Lishman, Jake and
    Gadhvi, Rushiraj and Agarwal, Rochisha and Galicia, Asier and Shammah, Nathan and
    Nation, Paul and Johansson, J. R. and Ahmed, Shahnawaz and Cross, Simon and
    Pitchford, Alexander and Nori, Franco
  },
  journal = {Physics Reports},
  volume = {1153},
  pages = {1-62},
  year = {2026},
  issn = {0370-1573},
  doi = {10.1016/j.physrep.2025.10.001},
  url = {https://www.sciencedirect.com/science/article/pii/S0370157325002704},
}

@article{fishman2022itensor,
  title={The ITensor software library for tensor network calculations},
  author={Fishman, Matthew and White, Steven and Stoudenmire, Edwin Miles},
  journal={SciPost Physics Codebases},
  pages={004},
  year={2022}
}

@article{Zhou-natcomm18,
  title={Achieving the Heisenberg limit in quantum metrology using quantum error correction},
  author={Zhou, Sisi and Zhang, Mengzhen and Preskill, John and Jiang, Liang},
  journal={Nature communications},
  volume={9},
  number={1},
  pages={78},
  year={2018},
  publisher={Nature Publishing Group UK London},
  url={https://www.nature.com/articles/s41467-017-02510-3}
}

@ARTICLE{11396347,
  author={Novak, Oskar and Rengaswamy, Narayanan},
  journal={IEEE Transactions on Quantum Engineering}, 
  title={Explaining Robust Quantum Metrology by Counting Codewords}, 
  year={2026},
  volume={7},
  number={},
  pages={1-12},
  doi={10.1109/TQE.2026.3664680}}

@article{manzano2020short,
  title={A short introduction to the Lindblad master equation},
  author={Manzano, Daniel},
  journal={Aip advances},
  volume={10},
  number={2},
  year={2020},
  publisher={AIP Publishing}
}

@article{Banchi_2021,
   title={Measuring Analytic Gradients of General Quantum Evolution with the Stochastic Parameter Shift Rule},
   volume={5},
   ISSN={2521-327X},
   url={http://dx.doi.org/10.22331/q-2021-01-25-386},
   DOI={10.22331/q-2021-01-25-386},
   journal={Quantum},
   publisher={Verein zur Forderung des Open Access Publizierens in den Quantenwissenschaften},
   author={Banchi, Leonardo and Crooks, Gavin E.},
   year={2021},
   month=jan, pages={386} }

@misc{kingma2017adammethodstochasticoptimization,
      title={Adam: A Method for Stochastic Optimization}, 
      author={Diederik P. Kingma and Jimmy Ba},
      year={2017},
      eprint={1412.6980},
      archivePrefix={arXiv},
      primaryClass={cs.LG},
      url={https://arxiv.org/abs/1412.6980}, 
}

@article{Larocca_2025,
   title={Barren plateaus in variational quantum computing},
   volume={7},
   ISSN={2522-5820},
   url={http://dx.doi.org/10.1038/s42254-025-00813-9},
   DOI={10.1038/s42254-025-00813-9},
   number={4},
   journal={Nature Reviews Physics},
   publisher={Springer Science and Business Media LLC},
   author={Larocca, Martín and Thanasilp, Supanut and Wang, Samson and Sharma, Kunal and Biamonte, Jacob and Coles, Patrick J. and Cincio, Lukasz and McClean, Jarrod R. and Holmes, Zoë and Cerezo, M.},
   year={2025},
   month=mar, pages={174–189} }

@article{PhysRevLett.91.147902,
  title = {Efficient Classical Simulation of Slightly Entangled Quantum Computations},
  author = {Vidal, Guifr\'e},
  journal = {Phys. Rev. Lett.},
  volume = {91},
  issue = {14},
  pages = {147902},
  numpages = {4},
  year = {2003},
  month = {Oct},
  publisher = {American Physical Society},
  doi = {10.1103/PhysRevLett.91.147902},
  url = {https://link.aps.org/doi/10.1103/PhysRevLett.91.147902}
}

@article{PhysRevLett.107.070601,
  title = {Time-Dependent Variational Principle for Quantum Lattices},
  author = {Haegeman, Jutho and Cirac, J. Ignacio and Osborne, Tobias J. and Pi\ifmmode \check{z}\else \v{z}\fi{}orn, Iztok and Verschelde, Henri and Verstraete, Frank},
  journal = {Phys. Rev. Lett.},
  volume = {107},
  issue = {7},
  pages = {070601},
  numpages = {5},
  year = {2011},
  month = {Aug},
  publisher = {American Physical Society},
  doi = {10.1103/PhysRevLett.107.070601},
  url = {https://link.aps.org/doi/10.1103/PhysRevLett.107.070601}
}

@misc{westhoff2025tensornetworkframeworklindbladian,
      title={A Tensor Network Framework for Lindbladian Spectra and Steady States}, 
      author={Philipp Westhoff and Mattia Moroder and Ulrich Schollwöck and Sebastian Paeckel},
      year={2025},
      eprint={2509.07709},
      archivePrefix={arXiv},
      primaryClass={quant-ph},
      url={https://arxiv.org/abs/2509.07709}, 
}

@article{PhysRevLett.87.167902,
  title = {Quantum Fingerprinting},
  author = {Buhrman, Harry and Cleve, Richard and Watrous, John and de Wolf, Ronald},
  journal = {Phys. Rev. Lett.},
  volume = {87},
  issue = {16},
  pages = {167902},
  numpages = {4},
  year = {2001},
  month = {Sep},
  publisher = {American Physical Society},
  doi = {10.1103/PhysRevLett.87.167902},
  url = {https://link.aps.org/doi/10.1103/PhysRevLett.87.167902}
}

@article{ramsey1950molecular,
  title={A molecular beam resonance method with separated oscillating fields},
  author={Ramsey, Norman F},
  journal={Physical Review},
  volume={78},
  number={6},
  pages={695},
  year={1950},
  publisher={APS}
}

@article{Ouyang-pra23,
  title = {Describing quantum metrology with erasure errors using weight distributions of classical codes},
  author = {Ouyang, Yingkai and Rengaswamy, Narayanan},
  journal = {Phys. Rev. A},
  volume = {107},
  issue = {2},
  pages = {022620},
  numpages = {16},
  year = {2023},
  month = {Feb},
  publisher = {American Physical Society},
  doi = {10.1103/PhysRevA.107.022620},
  url = {https://link.aps.org/doi/10.1103/PhysRevA.107.022620}
}

@article{wiebe2014quantum,
  title={Quantum algorithms for nearest-neighbor methods for supervised and unsupervised learning},
  author={Wiebe, Nathan and Kapoor, Ashish and Svore, Krysta},
  journal={arXiv preprint arXiv:1401.2142},
  year={2014}
}

@article{PhysRevLett.124.060503,
  title = {Beyond the Swap Test: Optimal Estimation of Quantum State Overlap},
  author = {Fanizza, M. and Rosati, M. and Skotiniotis, M. and Calsamiglia, J. and Giovannetti, V.},
  journal = {Phys. Rev. Lett.},
  volume = {124},
  issue = {6},
  pages = {060503},
  numpages = {6},
  year = {2020},
  month = {Feb},
  publisher = {American Physical Society},
  doi = {10.1103/PhysRevLett.124.060503},
  url = {https://link.aps.org/doi/10.1103/PhysRevLett.124.060503}
}

@InProceedings{fang_et_al:LIPIcs.ESA.2025.4,
  author =	{Fang, Wang and Wang, Qisheng},
  title =	{{Optimal Quantum Algorithm for Estimating Fidelity to a Pure State}},
  booktitle =	{33rd Annual European Symposium on Algorithms (ESA 2025)},
  pages =	{4:1--4:12},
  series =	{Leibniz International Proceedings in Informatics (LIPIcs)},
  ISBN =	{978-3-95977-395-9},
  ISSN =	{1868-8969},
  year =	{2025},
  volume =	{351},
  editor =	{Benoit, Anne and Kaplan, Haim and Wild, Sebastian and Herman, Grzegorz},
  publisher =	{Schloss Dagstuhl -- Leibniz-Zentrum f{\"u}r Informatik},
  address =	{Dagstuhl, Germany},
  URL =		{https://drops.dagstuhl.de/entities/document/10.4230/LIPIcs.ESA.2025.4},
  URN =		{urn:nbn:de:0030-drops-244727},
  doi =		{10.4230/LIPIcs.ESA.2025.4}
}

@article{Braunstein-prl94,
  title = {Statistical distance and the geometry of quantum states},
  author = {Braunstein, Samuel L. and Caves, Carlton M.},
  journal = {Phys. Rev. Lett.},
  volume = {72},
  issue = {22},
  pages = {3439--3443},
  numpages = {0},
  year = {1994},
  month = {May},
  publisher = {American Physical Society},
  doi = {10.1103/PhysRevLett.72.3439},
  url = {https://link.aps.org/doi/10.1103/PhysRevLett.72.3439}
}

@article{le2023variational,
  title={Variational quantum metrology for multiparameter estimation under dephasing noise},
  author={Le, Trung Kien and Nguyen, Hung Q and Ho, Le Bin},
  journal={Scientific Reports},
  volume={13},
  number={1},
  pages={17775},
  year={2023},
  publisher={Nature Publishing Group UK London}
}

@article{meyer2021variational,
  title={A variational toolbox for quantum multi-parameter estimation},
  author={Meyer, Johannes Jakob and Borregaard, Johannes and Eisert, Jens},
  journal={npj Quantum Information},
  volume={7},
  number={1},
  pages={89},
  year={2021},
  publisher={Nature Publishing Group UK London}
}

@article{koczor2020variational,
  title={Variational-state quantum metrology},
  author={Koczor, B{\'a}lint and Endo, Suguru and Jones, Tyson and Matsuzaki, Yuichiro and Benjamin, Simon C},
  journal={New Journal of Physics},
  volume={22},
  number={8},
  pages={083038},
  year={2020},
  publisher={IOP Publishing}
}

@misc{allen2025quantumcomputingenhancedsensing,
      title={Quantum Computing Enhanced Sensing}, 
      author={Richard R. Allen and Francisco Machado and Isaac L. Chuang and Hsin-Yuan Huang and Soonwon Choi},
      year={2025},
      eprint={2501.07625},
      archivePrefix={arXiv},
      primaryClass={quant-ph},
      url={https://arxiv.org/abs/2501.07625}, 
}

@misc{xu2025learningrestoreheisenberglimit,
      title={Learning to Restore Heisenberg Limit in Noisy Quantum Sensing via Quantum Digital Twin}, 
      author={Hang Xu and Tailong Xiao and Jingzheng Huang and Jianping Fan and Guihua Zeng},
      year={2025},
      eprint={2508.11198},
      archivePrefix={arXiv},
      primaryClass={quant-ph},
      url={https://arxiv.org/abs/2508.11198}, 
}

@misc{marrero2026encodedquantumsignalprocessing,
      title={Encoded Quantum Signal Processing for Heisenberg-Limited Metrology}, 
      author={Carlos Ortiz Marrero and Rui Jie Tang and Nathan Wiebe},
      year={2026},
      eprint={2603.22798},
      archivePrefix={arXiv},
      primaryClass={quant-ph},
      url={https://arxiv.org/abs/2603.22798}, 
}

@misc{kwon2026restoringheisenbergscalingtime,
      title={Restoring Heisenberg scaling in time via autonomous quantum error correction}, 
      author={Hyukgun Kwon and Uwe R. Fischer and Seung-Woo Lee and Liang Jiang},
      year={2026},
      eprint={2504.13168},
      archivePrefix={arXiv},
      primaryClass={quant-ph},
      url={https://arxiv.org/abs/2504.13168}, 
}

@article{Wang_2020,
   title={Prospect of using Grover’s search in the noisy-intermediate-scale quantum-computer era},
   volume={102},
   ISSN={2469-9934},
   url={http://dx.doi.org/10.1103/PhysRevA.102.042609},
   DOI={10.1103/physreva.102.042609},
   number={4},
   journal={Physical Review A},
   publisher={American Physical Society (APS)},
   author={Wang, Yulun and Krstic, Predrag S.},
   year={2020},
   month=oct }

@inbook{Escudero_2023,
   title={Assessing the Impact of Noise on Quantum Neural Networks: An Experimental Analysis},
   ISBN={9783031407253},
   ISSN={1611-3349},
   url={http://dx.doi.org/10.1007/978-3-031-40725-3_27},
   DOI={10.1007/978-3-031-40725-3_27},
   booktitle={Hybrid Artificial Intelligent Systems},
   publisher={Springer Nature Switzerland},
   author={Escudero, Erik Terres and Alamo, Danel Arias and Gómez, Oier Mentxaka and Bringas, Pablo García},
   year={2023},
   pages={314–325} }

@article{zhang2021distributed,
  title={Distributed quantum sensing},
  author={Zhang, Zheshen and Zhuang, Quntao},
  journal={Quantum Science \& Technology},
  volume={6},
  number={4},
  pages={043001},
  year={2021},
  publisher={IOP Publishing}
}

@misc{wang2025noiseresilientquantummetrologyquantum,
      title={Noise-Resilient Quantum Metrology with Quantum Computing}, 
      author={Xiangyu Wang and Chenrong Liu and Xue Lin and Yu Tian and Yishan Li and Xinfang Nie and Yufang Feng and Yuxuan Zheng and Ying Dong and Xinqing Wang and Dawei Lu},
      year={2025},
      eprint={2509.00771},
      archivePrefix={arXiv},
      primaryClass={quant-ph},
      url={https://arxiv.org/abs/2509.00771}, 
}
\bibliographystyle{apsrev}

\end{document}